\def\msol{M_{\odot}}

\def\mh1{{M_{\hbox{\rs HI}}}}
\def\21cm{{21 \hskip -2pt cm}}

\def\hubunits{km \hskip -2pt s$^{-1}$ \hskip -2pt Mpc$^{-1}$}

\font\smcap=cmcsc10

\def\h1{{\smcap HI}}

\documentstyle[12pt,aaspp4]{article}

\newcommand{\etal}{et al.}
\newcommand{\tsph}{TreeSPH}
\newcommand{\lya}{Ly$\alpha$\ }

\def\cf{{\it cf.\/} }
\def\be{\begin{equation}}
\def\ee{\end{equation}}
\def\bdm{\begin{displaymath}}
\def\edm{\end{displaymath}}

\def\kmpersec{km s$^{-1}$}
\def\NHI{N_{\rm HI}}

\newbox\grsign \setbox\grsign=\hbox{$>$} \newdimen\grdimen \grdimen=\ht\grsign
\newbox\simlessbox \newbox\simgreatbox
\setbox\simgreatbox=\hbox{\raise.5ex\hbox{$>$}\llap
     {\lower.5ex\hbox{$\sim$}}}\ht1=\grdimen\dp1=0pt
\setbox\simlessbox=\hbox{\raise.5ex\hbox{$<$}\llap
     {\lower.5ex\hbox{$\sim$}}}\ht2=\grdimen\dp2=0pt

\newcommand{\simlt}{\mathrel{\copy\simlessbox}}

\begin{document}

\title{TESTING COSMOLOGICAL MODELS AGAINST THE ABUNDANCE OF DAMPED 
LYMAN-ALPHA ABSORBERS}

\author{
Jeffrey P. Gardner\altaffilmark{1},
Neal Katz\altaffilmark{1,2},
David H. Weinberg\altaffilmark{3},
Lars Hernquist\altaffilmark{4,5}
}
\affil{E-mail:  gardner@astro.washington.edu, nsk@astro.washington.edu,
 lars@helios.ucsc.edu, dhw@payne.mps.ohio-state.edu}
\altaffiltext{1}{University of Washington, Department of Astronomy, 
Seattle, WA 98195}
\altaffiltext{2}{University of Massachusetts, 
Department of Physics and Astronomy, Amherst, MA 01003-4525}
\altaffiltext{3}{Ohio State University, 
Department of Astronomy, Columbus, OH 43210}
\altaffiltext{4}{University of California, 
Lick Observatory, Santa Cruz, CA 95064}
\altaffiltext{5}{Presidential Faculty Fellow}

\begin{abstract}

We calculate the number of damped \lya absorbers expected in various
popular cosmological models as a function of
redshift and compare
our predictions with observed abundances.  The Press-Schechter
formalism is used to obtain the distribution of halos with circular
velocity in different cosmologies, 
and we calibrate the relation between circular velocity and
absorption cross-section using detailed gas dynamical simulations
of a ``standard'' cold dark matter (CDM) model.
Because of this calibration,
our approach makes more realistic assumptions about the absorption
properties of collapsed objects than previous, analytic calculations
of the damped \lya abundance.
CDM models with $\Omega_0=1$, $H_0=50\;$\hubunits,
baryon density $\Omega_b=0.05$, and scale-invariant primeval fluctuations
reproduce the observed incidence and redshift evolution of damped \lya 
absorption to within observational uncertainty, for both COBE 
normalization ($\sigma_8=1.2$) and a lower normalization ($\sigma_8=0.7$)
that better matches the observed cluster abundance at $z=0$.
A tilted ($n=0.8$, $\sigma_8=0.7$) CDM model tends to underproduce
absorption, especially at $z=4$.  With COBE normalization, a CDM model
with $\Omega_0=0.4$, $\Omega_\Lambda=0.6$ gives an acceptable fit to the
observed absorption; an open CDM model is marginally acceptable if
$\Omega_0 \geq 0.4$ and strongly inconsistent with the $z=4$ data if
$\Omega_0=0.3$.  Mixed dark matter models tend not to
produce sufficient absorption, being roughly comparable to tilted
CDM models if $\Omega_\nu = 0.2$ and failing drastically if
$\Omega_\nu = 0.3$.

\end{abstract}


\section{Introduction}

Damped \lya (DLA) absorbers --- hydrogen absorption systems with neutral
column density $\NHI \geq 10^{20.3}\;{\rm cm}^{-2}$ ---
are important tracers of high density structure in the early universe.
They are far more common than quasars or luminous radio galaxies, 
and various lines of evidence, both coincidences of numbers and
properties of observed systems, indicate that DLA systems are the
high-redshift progenitors of ``normal'' present-day galaxies
(see, e.g., Wolfe \etal\ 1995;
Djorgovski \etal\ 1996; Fontana \etal\ 1996; 
Prochaska \& Wolfe 1997).
In this {\it Letter}, we use a semi-analytic method,
calibrated against numerical simulations, to test the ability of current 
theories of structure formation to reproduce the observed abundance
of DLA absorbers at $2 \leq z \leq 4$.

The first attempts to employ DLA absorption as a test of
cosmological scenarios used the Press-Schechter (1974) formalism
to predict the abundance of dark matter halos and adopted simplifying,
usually conservative assumptions about the amount of gas within
these halos that would cool and become neutral
(Kauffmann \& Charlot 1994; Mo \& Miralda-Escud\'e 1994;
for recent analyses along similar lines see
Liddle \etal\ 1996ab).
Ma \& Bertschinger (1994) and Klypin \etal\ (1995) 
proceeded in a similar fashion, but computed
the halo mass function from dissipationless N-body simulations.  
Kauffmann (1996) presented a more sophisticated semi-analytic treatment
of DLA systems and disk galaxy formation in the standard cold dark 
matter (CDM) model, incorporating more realistic assumptions about
the cooling and settling of gas within halos and subsequent
star formation.

A fundamentally different approach was taken by Katz \etal\ 
(1996; hereafter KWHM), who used numerical 
simulations incorporating dark
matter and gas in the presence of a photoionizing background to show
that damped \lya systems originate naturally from initial conditions
similar to those which are believed to have produced the observed
large-scale structure.  In the KWHM simulations, damped absorption
results from concentrations of dense gas embedded in more extended and
more massive dark matter halos.  The masses and abundances of these
objects are comparable to those of galaxies, supporting the view that
damped \lya systems are a natural by-product of galaxy formation.
Lower column density, \lya forest absorption arises in more diffuse,
highly photoionized structures, and the combined results of KWHM and
Hernquist \etal\ (1996) show that a simulation of the standard CDM model
reproduces the observed distribution of neutral hydrogen column densities
quite well over most of the range 
$10^{14}\;{\rm cm}^{-2} \leq \NHI \leq 10^{22}\;{\rm cm}^{-2}$.

Gardner \etal\ (1997; hereafter GKHW) improved on KWHM by using a
semi-analytic method to correct for damped absorption in halos
below the KWHM resolution limit.  Using KWHM's simulation and
higher resolution simulations focused on individual, collapsing
halos, GKHW compute the mean projected area $\alpha(v_c,z)$
over which a halo of circular velocity $v_c$ produces 
damped \lya absorption
at redshift $z$.  They convolve this function with the Press-Schechter
formula for the abundance of halos as a function of $v_c$ to obtain
the total incidence of damped absorption in all halos.
The correction for unresolved objects raises the KWHM predictions
by roughly a factor of two, bringing them into good agreement with the
observed abundance of damped absorbers, though the predicted number
of Lyman limit systems remains low by about a factor of three.
GKHW also show that star formation (with the algorithm of 
Katz, Weinberg, \& Hernquist 1996) has little effect on the predicted
number of damped systems.

The method adopted in KWHM and GKHW can be applied to other cosmological
scenarios, and we believe that numerical simulations with gas dynamics
will eventually allow the most robust confrontations between predicted
and observed properties of DLA absorbers.  However, the fully numerical
approach is computationally very expensive, and it will be some time
before it can be applied to a broad parameter space of cosmological models.
This seems like an opportune moment to revisit the semi-analytic
approach for several reasons.  First, the theoretical ground has
shifted somewhat since the first investigations, prompted in part by
improved analyses of the COBE data (and the 4-year COBE data themselves)
and in part by further studies of the theoretical models, especially
the emergence of open-bubble inflation models as a viable and predictive 
setting for structure formation (e.g., Ratra \& Peebles 1994;
Bucher, Goldhaber, \& Turok 1995; Yamamoto, Sasaki, \& Tanaka 1995).
Second, new data have improved the observational constraints,
especially at $z \approx 4$ (Wolfe \etal\ 1995; Storrie-Lombardi \etal\ 1996).
Third, and most significant from the point of view of this paper, 
the GKHW results provide a way to calibrate the most uncertain element
of the semi-analytic method, the relation between halo circular velocity
and the cross-section for damped absorption.  It seems plausible that
this relation is driven primarily by ``local'' physics within the halo,
and for this paper we will therefore assume that the function 
$\alpha(v_c,z)$ found by GKHW for the standard CDM model also applies
to other cosmological scenarios.  With this assumption --- uncertain,
but probably the best approximation available until alternative models
are investigated numerically --- we can use the GKHW method to compute
the abundance of DLA absorbers in any specified cosmology with 
Gaussian initial conditions.  We describe this method in more detail
in \S 2, then present our results and discuss their implications in \S 3.

\section{Method}
\label{secMethod}

KWHM used \tsph\ (Hernquist \& Katz 1989) to evolve to redshift $z=2$
a periodic cube 22.22 comoving Mpc on a side containing dark matter and gas. 
The initial conditions were drawn from a CDM power spectrum with
$\sigma_8=0.7$, $\Omega_0=1$, $h=0.5$\footnote[1]{Throughout,
$h\equiv H_0/(100\;$\hubunits), where $H_0$ is Hubble's constant today.},
and $\Omega_b=0.05$, where
$\sigma_8$ is the present-day rms mass fluctuation within a sphere of
radius 8 $h^{-1}$ Mpc, and $\Omega_b$ and $\Omega_0$ are the present-day
fractions of closure density in the form of baryons and matter
respectively.  The simulation included a spatially uniform background
radiation field of
intensity $J(\nu)=J_0(\nu_0/\nu)F(z)$, where $\nu_0$ is the
Lyman-limit frequency, $J_0=10^{-22}$ erg s$^{-1}$ cm$^{-2}$ sr$^{-1}$
Hz$^{-1}$, and $F(z)=0$ if $z>6$, $F(z)=4/(1+z)$ if $3 \le z \le 6$,
and $F(z)=1$ if $2<z<3$.  More detailed descriptions of the
simulation and methods can be found in GKHW, KWHM, and Katz \etal\ (1996).

Given their limited particle number, KWHM were able to resolve halos
only down to a mass corresponding to a virial circular velocity of
$v_c \approx 100$ \kmpersec\ (about $10^{11} \msol$ at $z=2$; see,
also Weinberg, Hernquist, \& Katz 1997).  Since
Quinn, Katz, \& Efstathiou (1996) find that halos with $v_c>37$
\kmpersec\ are capable of hosting damped \lya absorbers, it is clear
that KWHM's simulation systematically underestimates the true prediction
of this CDM model for the incidence of damped absorption.
GKHW combine the background radiation field of KWHM with the initial
conditions of Quinn \etal\ (1996) to simulate small regions with
sufficient detail to resolve halos with circular velocities
below 50 \kmpersec.  The high resolution volumes are
too small to calculate the number of damped systems (they were chosen
to lie in regions where small halos would form; see Quinn \etal\ 1996), but
by using the halos from both the high resolution and KWHM simulations, GKHW
estimate $\alpha(v_c,z)$, the cross-section for a halo of circular
velocity $v_c$ to produce damped \lya absorption at redshift $z$.  
GKHW show that $\alpha(v_c,z)$ can be fitted by a power law in $v_c$, 
with an exponential cutoff
to reproduce the absence of absorbers in halos with $v_c < 37$
\kmpersec\ (Quinn \etal\ 1996).  The fitting procedure is described in
GKHW, which also lists parameters of the fitted relation at $z=2$, 3, and 4. 

According to Press-Schechter theory, the number density $N(M,z)$ of
collapsed objects in the mass range $M\longrightarrow M+dM$ at redshift $z$ is
\be
        N(M,z) dM = \sqrt{2\over \pi} {\rho_0\over M} 
         {\delta_c\over \sigma_0} \left({\gamma R_f\over R_*}\right)^2
         \exp{\left({-\delta_c^2\over 2\sigma_0^2}\right)} dM ,
\label{PSNofM}
\end{equation}
where $\rho_0$ is the mean comoving mass density and $R_f$ is the Gaussian
filter radius given by $M= (2\pi)^{3/2} \rho_0 R_f^3$.  The parameters
$\sigma_0$, $\gamma$ and $R_*$ are related to moments of the power
spectrum and are defined in Bardeen \etal\ (1986).  As detailed in GKHW,
multiplying $N(M,z)$ by $\alpha(v_c,z)$ and integrating from $v_c$ to
infinity yields the number of damped absorbers per unit redshift residing
in halos of circular velocity greater than $v_c$:
\be
        n(z,v_c)= {dr \over dz} \int_{M(v_c)}^{\infty} N(M',z)
                \alpha(v_c',z) dM' ,
\label{nofzM}
\ee
where $v_c'$ is the circular velocity at the virial radius
of a halo of virial mass $M'$, and $r$ is comoving distance.
With $v_c=37$ \kmpersec, the lower limit for damped absorption
found by Quinn \etal\ (1996), equation~(\ref{nofzM}) yields the
total incidence of DLA absorption $n(z)$,
defined to be the number of absorption systems per unit redshift
with $\NHI \geq 10^{20.3}\;{\rm cm}^{-2}$.
In this paper, we adopt the $\alpha(v_c,z)$ relations found by GKHW
for the standard CDM model but combine them with the halo
abundance $N(M,z)$ predicted for other cosmological scenarios.

For all the Press-Schechter predictions we use a Gaussian filter with
$\delta_c = 1.69$.  This $\delta_c$ is higher than that
conventionally used with a Gaussian filter because of the way that we
identify halos and assign them circular velocities (see GKHW),
which we believe most closely corresponds to the conception 
of halos in the Press-Schechter formalism.
As explained in GKHW, this value of $\delta_c$ fits the 
numerically derived halo mass function of KWHM quite well. 
Our full procedure correctly predicts $n(z,v_c)$ 
in the KWHM simulations for damped systems in halos above the
$v_c=100$ \kmpersec\ resolution limit.  In our calculations of
$N(M,z)$, we use the transfer functions of
Bardeen \etal\ (1986) for cold dark matter models and Ma (1996) 
for mixed dark matter models.

\section{Results}
\label{secResults}

\begin{figure}
\epsfxsize=6.5truein
\centerline{\epsfbox[18 144 590 717]{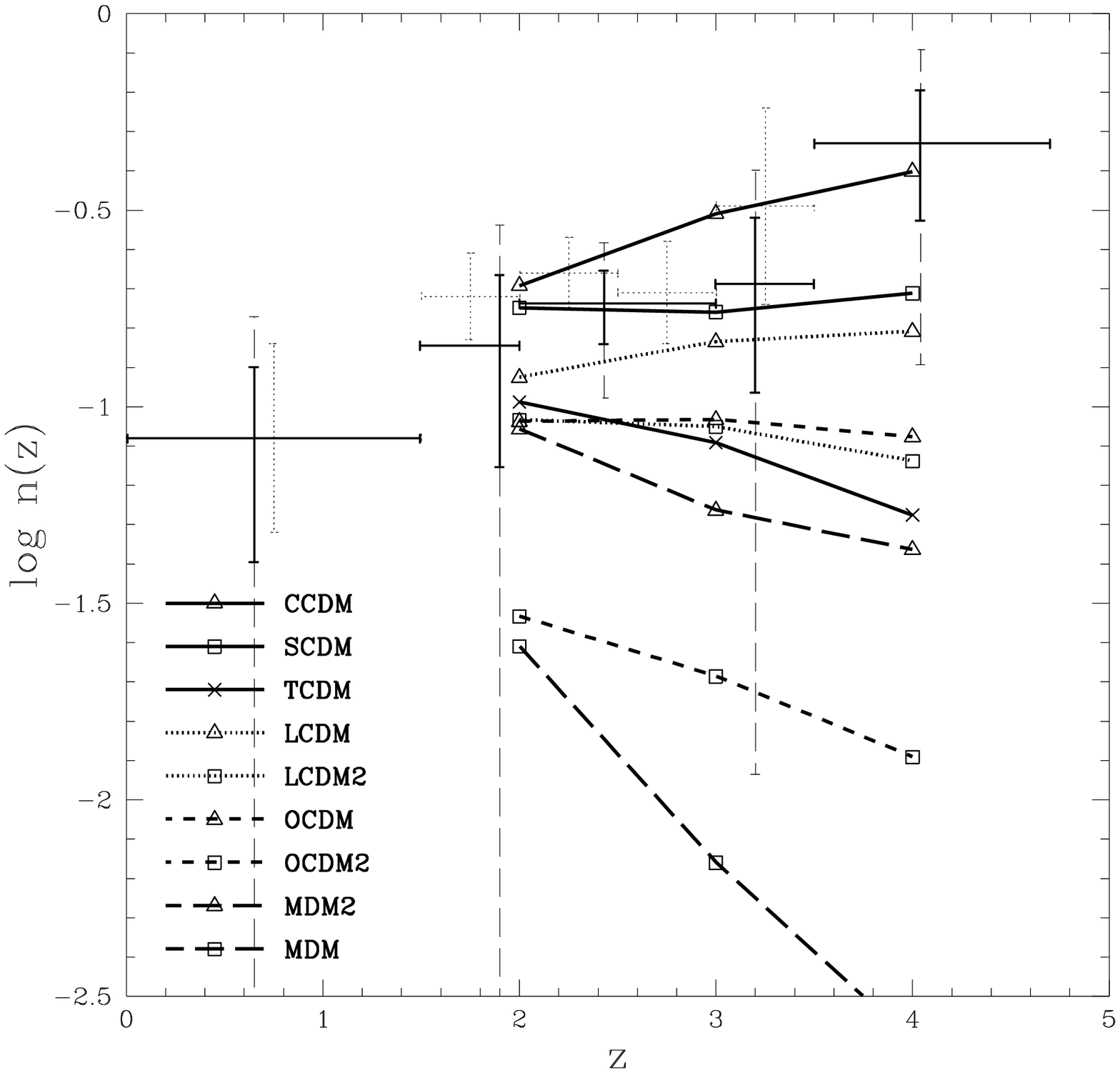}}
\caption{
\label{figDLA}
The predicted number $n(z)$ of DLA systems per unit redshift
for various cosmological models vs. observational data.  
The solid error crosses ($1\sigma$)
are reproduced from Storrie-Lombardi \etal\ (1996), with the vertical
dashed error bars indicating $2\sigma$ error.  The dotted error crosses 
($1\sigma$) are taken from Wolfe \etal\ (1995).  The heavy curves connecting
$z=2,3,4$ are values calculated for each model.  The model names
correspond to the names in Table~\protect\ref{tabDLA}, where their
parameters are listed.}
\end{figure}

\begin{figure}
\epsfxsize=6.5truein
\centerline{\epsfbox[18 144 590 718]{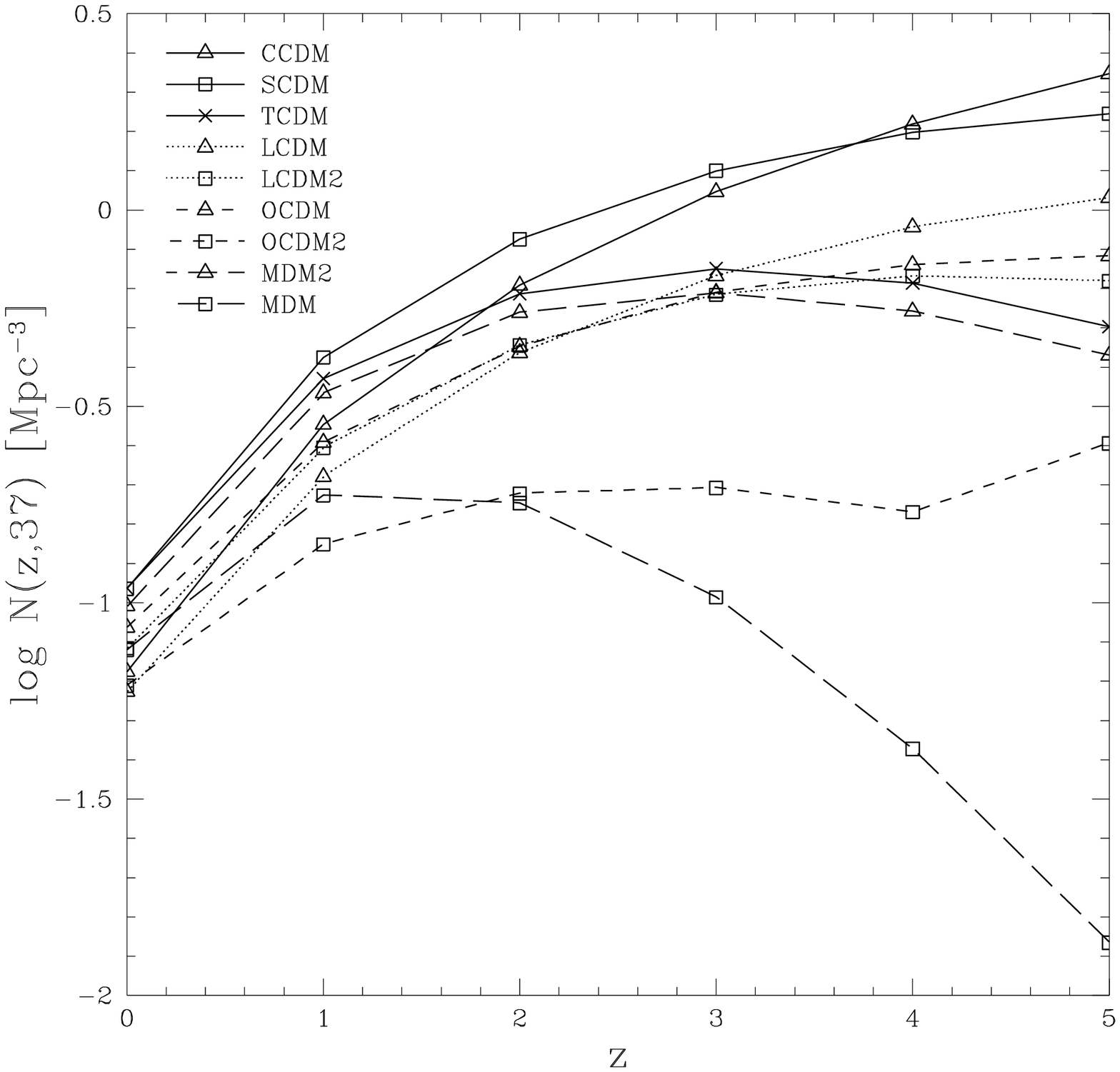}}
\caption{Comoving number density of halos with $v_c>37$ \kmpersec,
$N(z,37)$, vs.\ redshift.  The models and labeling are as in
Figure~\protect\ref{figDLA}.}
\label{figNplot}
\end{figure}

Table~\ref{tabDLA} compares the predicted numbers of DLA absorbers
to the observational results reported by Storrie-Lombardi \etal\ (1996).
These results are illustrated in Figure~\ref{figDLA},  
which also includes the observational data of Wolfe \etal\ (1995).
To aid the interpretation of this Figure, we show in Figure~\ref{figNplot}
the comoving number density of halos with $v_c>37$ \kmpersec\
obtained by integrating equation~(\ref{PSNofM}) for each model.
Starting at high redshift, the halo abundance first rises as rms
fluctuations on the $v_c \sim 37$ \kmpersec\ mass scale go nonlinear,
then declines as the first generation of halos above this threshold
merges into a smaller number of more massive halos.  For some of
our models, this decline has already begun by $z=5$.

\begin{table}
\begin{tabular}{lllllllll}
 \tableline\tableline
\multicolumn{6}{l}{ } & \multicolumn{3}{c}{$n(z)$} \\ \cline{7-9}
Name & $\sigma_8$ & $\Omega_0$ & $\Omega_{\Lambda}$ & $h$ & Other &
$z=2$ & $z=3$ & $z=4$ \\ \tableline

CCDM\tablenotemark{a}&   1.2&    1&      0&      0.5&    &
        0.203&      0.310&   0.397 \\
SCDM&   0.7&    1&      0&      0.5&    &
        0.179&      0.174&   0.194 \\
TCDM\tablenotemark{b}& 0.7&    1&      0&      0.5&
$n=0.8$&
        0.103&         0.0811&      0.0530 \\
LCDM\tablenotemark{a}&   1.1&   0.4&    0.6&    0.65&   &
        0.119&       0.146&      0.155 \\
LCDM2\tablenotemark{c}&  0.79&    0.4&    0.6&    0.65&   &
        0.0926&      0.0889&     0.0730 \\
OCDM\tablenotemark{d}&  0.65&   0.4&    0&      0.65&   &
        0.0916&      0.0926&     0.0836 \\
OCDM2\tablenotemark{d}& 0.45&    0.3&    0&      0.65&   &
        0.0293&     0.0206&     0.0128 \\
MDM\tablenotemark{e}&   0.782&  1&      0&      0.5&    $\Omega_{\nu}=0.3$&
        0.0246&    0.00692&   0.00243 \\
MDM2\tablenotemark{e}&  0.808&  1&      0&      0.5&    $\Omega_{\nu}=0.2$&
        0.0875&    0.0545&    0.0432 \\
& & & & & & & & \\
& & & &\multicolumn{2}{c}{$1.5<z<2.0$}&\multicolumn{1}{c}{$2.0<z<3.0$}&
\multicolumn{1}{c}{$3.0<z<3.5$}&\multicolumn{1}{c}{$3.5<z<4.7$} \\ \cline{5-9}
Observed \tablenotemark{f}& &  & &
\multicolumn{2}{c}{$0.14\pm 0.073$}& $0.18\pm 0.039$&
        $0.21\pm 0.097$&        $0.47\pm 0.17$ \\ \tableline\tableline

\end{tabular}

\caption{The predicted abundance of DLA systems, $n(z)$, 
for a variety of cosmological models.  
$\Omega_\Lambda \equiv \Lambda/(3 H_0^2)$ where $\Lambda$ 
is the cosmological constant.
References for model normalizations are listed below.
The final row shows observed data and $1\sigma$ errors.
}
\tablenotetext{a}{Bunn \& White (1996)}
\tablenotetext{b}{White \etal\ (1996)}
\tablenotetext{c}{Cen \etal\ (1994), Kofman, Gnedin, \& Bahcall (1993)}
\tablenotetext{d}{Gorksi \etal\ (1996)}
\tablenotetext{e}{Ma (1996)}
\tablenotetext{f}{Storrie-Lombardi \etal\ (1996)}
\label{tabDLA}
\end{table}


The model producing the most DLA absorption and the best agreement
with the $z=4$ data is CCDM: 
COBE-normalized, $\Omega_0=1$ CDM with
scale-invariant ($n=1$) primeval fluctuations and $h=0.5$.  
However, this model is known to produce excessively
massive clusters at $z=0$.  SCDM, the ``standard'' CDM model
used in GKHW, has an identical power spectrum shape but is normalized
to produce more reasonable (still somewhat high) cluster masses,
while disagreeing with the COBE results.  In this case, $n(z)$ is lower due
to the decrease in power (\cf Figure~\ref{figNplot}), but the model still
matches the DLA observations within current uncertainties.  
It is possible to construct an $\Omega_0=1$ CDM model that simultaneously
fits cluster masses and COBE data by
further adjustments, such as tilting the primeval power spectrum
to $n<1$ and raising the baryon fraction.  White \etal\ (1996) give an
example of such a model, which we designate TCDM: 
with $n=0.8$ and $\Omega_b=0.1$, COBE normalization implies 
$\sigma_8=0.7$ just as for our SCDM model.  However, the tilt and 
high $\Omega_b$ reduce the amount of small-scale
power relative to SCDM, reducing the halo abundance at
$z>2$ (Figure~\ref{figNplot}) and producing an $n(z)$ significantly
below the Storrie-Lombardi \etal\ (1996) data for $z=3$
and $z=4$.  There are other combinations of $n$, $\Omega_b$, and $h$
that can match COBE and still have CDM, $\Omega_0=1$, and $\sigma_8 \sim 0.7$,
but since the necessary changes all reduce the amount of power on small
scales, the problem shown here is likely to be generic to all such models.

It is important to note that TCDM is the only model we analyze that
has $\Omega_b h^2$ different from $0.0125$, and we
have not allowed the increased baryon density to change the DLA
cross-section in our treatment here.  It is not difficult to imagine
that the increase in baryon density could alter the absorption
produced in high-redshift halos.  Larger $\Omega_b$ implies an increased gas
supply for a halo of a given mass, which could simply lead to a higher
incidence of damped absorption.  On the other hand, it could also lead to more
rapid cooling and more dense concentrations of cold collapsed gas,
thereby decreasing $n(z)$.  Consequently, our results for this model
are not robust enough to rule it out until the influence of $\Omega_b$
on $n(z)$ has been more thoroughly investigated.

CDM models with a cosmological constant retain the flat universe
preferred by inflationary models while decreasing the density of matter,
permitting more reasonable cluster masses with a COBE normalization.
They also match the shape of the power spectrum on galaxy cluster scales
more closely than SCDM, and they avoid the severe problems of $\Omega_0=1$
models concerning the age of the universe and the cluster baryon
fraction.  Our LCDM model, COBE-normalized with $\Omega_0=0.4$ and
$h=0.65$, generally falls within the $1\sigma$ errors of the
Storrie-Lombardi data except at $z=4$.  Many of the systems which
contribute to the $z=4$ data point have not yet been confirmed by
high-resolution spectroscopy, however, so the value could decrease
with future observations.  Hence LCDM could be considered acceptable.
LCDM2 is an identical model with the lower normalization favored by
Cen \etal\ (1994).  The lessened power of this model leads to
a lower $n(z)$.  Actually achieving this low normalization while being
consistent with COBE would require a tilt or a lower $h$, which in turn
would reduce small scale power, further exacerbating the problem.

It is also possible to preserve an inflationary scenario in an open CDM
universe.  OCDM shows a COBE-normalized open model similar to LCDM
but with no cosmological constant.  The open model has a lower mass
fluctuation amplitude and hence a lower $n(z)$
than LCDM.  An open CDM model with $\Omega_0=0.5$,
$\sigma_8 \sim 0.9$ (not shown)
skirts the lower end of the observational data and might
be acceptable within the current uncertainties.
The amount of small scale power implied by 
COBE normalization in open models decreases rapidly with decreasing
$\Omega_0$ (\cf Figure~\ref{figNplot}).  For
$\Omega_0=0.3$ (OCDM2), $\sigma_8$ is reduced to near $0.5$, and
the predicted $n(z)$ is well below the observations.

The most popular alternative to lowering $\Omega_0$ is to add hot dark
matter.  This reduces $\sigma_8$ given by the COBE normalization
and makes the shape of the power spectrum closer to that
observed for galaxies.  The first incarnations of this mixed dark matter
model had
$\Omega_\nu=0.3$, which we reproduce here as MDM using the power
spectrum and normalization given in Ma (1996).  This model has strongly
suppressed structure formation at high redshift (\cf
Figure~\ref{figNplot}), and it drastically underproduces DLA
absorption.  Largely in recognition of this problem, $\Omega_\nu=0.2$,
shown here as MDM2, has become the favored version.  While more palatable
than MDM, this model still falls $2.8\sigma$ short for $2<z<3$ and
$2.5\sigma$ below observations at $z=4$.  This problem could possibly
be alleviated by further modifications such as ``anti-tilting''
the primeval spectrum to $n>1$ or
raising the Hubble constant to $h>0.5$, but these changes would raise
$\sigma_8$, which is already too high to yield a good match to cluster
abundances for $\Omega_0=1$ (White, Efstathiou, \& Frenk 1993;
Eke, Cole, \& Frenk 1996).
The most optimistic reading would be
that mixed-dark matter can barely squeak by the combined constraints
of cluster abundances on one hand and DLA's on the other.
Improvements in the observational data of DLA's and the firming up of
theoretical predictions by targeted simulations will probably close
this window of parameter space.

These results are consistent with the findings of Mo \&
Miralda-Escud\'e (1994), in which LCDM underproduces absorption at
$2 \simlt z \simlt 3$ by roughly a factor of three and
MDM (with $\Omega_\nu=0.3$)
fails in this range by many orders of magnitude.  Kauffmann \&
Charlot (1994) also conclude that standard CDM gives 
the best match to observations,
remarking upon the inability of MDM models to produce the necessary
cold, collapsed gas at higher redshifts.  Ma \& Bertschinger (1994)
find that an $\Omega_\nu=0.2$ MDM model achieves rough agreement with the
observations in the range $2.5<z<3.5$, but
they assume that the distribution of gas matches the distribution
of dark matter in the halos of their (dissipationless) simulation.
The KWHM simulations show that only a fraction of the gas in a virialized
halo cools and becomes neutral and that it is 
distributed primarily in one or more small, dense knots; both effects
can reduce the damped \lya absorption cross-section of a given halo.
Ma et al.\ (1997) have undertaken a new study of DLA absorption in the
mixed dark matter model, calibrating the Ma \& Bertschinger (1994)
dissipationless method against the KWHM simulation of SCDM, and they
also find that the $\Omega_\nu=0.2$ model underproduces the observed
DLA absorption.

The global cosmological parameters in the CCDM and
MDM models are the same as those in the SCDM model simulated by KWHM,
and we are therefore fairly confident that our method can accurately
calibrate their predictions of $n(z)$.
The TCDM, LCDM, and OCDM models, however,
adopt different values of $\Omega_b$, $\Omega_0$, $h$, and $\Lambda$,
altering the relations between overdensity and gas cooling rate and between
redshift and the age of the universe.  Our assumption that
$\alpha(v_c,z)$ is close to that of the SCDM model is therefore
more suspect.  GKHW find that the DLA cross-section at fixed $v_c$
decreases with time as the gas condenses into tighter configurations.
If we assumed a similar trend across changes of cosmological parameters,
then our predictions of $n(z)$ for TCDM, LCDM, and OCDM would
go down, so in this sense our conclusions about these models are
likely to be conservative.  However, more detailed numerical studies
will be required before we can build a definitive case against them.

As a concluding comment, we can generalize a point that we have already
made regarding the mixed dark matter models.  The model that produces
the best match to the high-redshift DLA data, CCDM, is grossly 
inconsistent with the mass function of galaxy clusters today.
The SCDM model, which fits the DLA data acceptably, is at best barely
compatible with the cluster data, which imply $\sigma_8 \approx 0.55$
for $\Omega_0=1$ rather than $\sigma_8=0.7$ (White \etal\ 1993;
Eke \etal\ 1996).  Our LCDM model skirts the low end of the DLA
data and the high end of the cluster data (see Cole \etal\ 1997).
Clearly these are specific instances of a general problem,
within the broad class of Gaussian/inflationary models considered here,
since most parameter changes that reduce cluster masses also reduce
DLA abundances.  Between the Scylla of the damped \lya
systems and the Charybdis of the cluster mass function lies a narrow
channel that few cosmological models will pass.

\acknowledgments

This work was supported in part by the Pittsburgh Supercomputing
Center, the National Center for Supercomputing Applications
(Illinois), the San Diego Supercomputing Center, NASA Theory Grants
NAGW-2422, NAGW-2523, NAG5-2882, and NAG5-3111, NASA HPCC/ESS Grant NAG 5-2213,
and the NSF under Grant ASC 93-18185 and the Presidential Faculty
Fellows Program.

\vfill\eject

\end{document}